\documentclass[12pt]{article}
\usepackage{axodraw}

\catcode`@=12
\begin{document}
\thispagestyle{empty}
\begin{flushright} 
UCRHEP-T347\\ 
September 2002\
\end{flushright}
\vspace{0.5in}
\begin{center}
{\LARGE	\bf Supersymmetric Axion-Neutrino Merger\\}
\vspace{1.5in}
{\bf Ernest Ma\\}
\vspace{0.2in}
{\sl Physics Department, University of California, Riverside, 
California 92521\\}
\vspace{2.0in}
\end{center}
\begin{abstract}\
The recently proposed supersymmetric $A_4$ model of the neutrino mass matrix 
is modified to merge with a previously proposed axionic solution of 
the strong CP problem.  The resulting model has only one input scale, 
i.e. that of $A_4$ symmetry breaking, which determines both the seesaw 
neutrino mass scale and the axion decay constant.  It also solves the 
$\mu$ problem and conserves $R$ parity automatically.
\end{abstract}
\newpage
\baselineskip 24pt

\section{Introduction}

There are a number of issues in particle physics which require theoretical 
clarification to go along with our present experimental knowledge.  As it 
stands, the minimal standard model (SM) has the following shortcomings: 
(1) absence of neutrino mass, (2) presence of the strong CP problem, and 
(3) presence of the hierarchy problem.  Supersymmetry is widely assumed to 
be the resolution of (3), but the minimal supersymmetric standard model 
(MSSM) still has no cure for (1) or (2).  Furthermore, it raises two other 
issues not present in the SM, namely (4) the nonconservation of baryon 
number and lepton number without the imposition of $R$ parity, and (5) the 
$\mu$ problem, i.e. why the allowed supersymmetric term $\mu \hat \phi_1 
\hat \phi_2$ in the superpotential has $\mu$ of the order of the supersymmetry 
breaking scale, instead of some other much larger scale such as the Planck 
scale or the string scale.

In this paper, two recent proposals \cite{ma01,bmv} are merged to form a 
cohesive theoretical understanding of all 5 issues.  The resulting 
supersymmetric model has only one input mass scale in the superpotential, 
which determines the anchor mass of the neutrino seesaw mechanism 
\cite{seesaw} as well as the scale of the spontaneous breaking of an 
anomalous global U(1) symmetry \cite{pq}. The latter solves the strong CP 
problem \cite{qcdcp} and predicts the existence of a very light pseudoscalar 
particle, the axion \cite{axi}.  Its specific implementation \cite{ma01} 
in this model also solves the $\mu$ problem and leads to the automatic 
conservation of $R$ parity.

\section{Description of Model}

The standard $SU(3)_C \times SU(2)_L \times U(1)_Y$ gauge symmetry is extended 
to include supersymmetry as well as an anomalous global $U(1)_{PQ}$ symmetry 
and the discrete non-Abelian symmetry $A_4$ \cite{a4}, which has 4 
irreducible representations, i.e. \underline {1}, \underline {1}$'$, 
\underline {1}$''$, and \underline {3}.  The usual superfields of the 
MSSM have the following assignments under $U(1)_{PQ}$ and $A_4$:
\begin{eqnarray}
&& \hat Q_i = (\hat u_i, \hat d_i), ~\hat L_i = (\hat \nu_i, \hat e_i) \sim 
(1/2,~\underline{3}), ~~~\hat \phi_{1,2} \sim (-1,~\underline{1}), \\ 
&& \hat u^c_1, ~\hat d^c_1, ~\hat e^c_1 \sim (-1/2,~\underline{1}), ~~~ 
\hat u^c_2, ~\hat d^c_2, ~\hat e^c_2 \sim (-1/2,~\underline{1}'), ~~~ 
\hat u^c_3, ~\hat d^c_3, ~\hat e^c_3 \sim (-1/2,~\underline{1}'').
\end{eqnarray}
The following quark, lepton, and Higgs superfields are then added:
\begin{eqnarray}
&& \hat U_i, ~\hat U^c_i, ~\hat D_i, ~\hat D^c_i, ~\hat E_i, ~\hat E^c_i, 
~\hat N^c_i \sim (1/2,~\underline{3}), \\ 
&& \hat \chi_i \sim (0,~\underline{3}), ~~~\hat \zeta_i \sim 
(-2,~\underline{3}),
~~~\hat S_1 \sim (-1,~\underline{1}), ~~~\hat S_2 \sim (2,~\underline{1}),
\end{eqnarray}
which are all $SU(2)_L$ singlets.  The superpotential of this model is then 
given by
\begin{eqnarray}
\hat W &=& {1 \over 2} M_\chi \hat \chi_i \hat \chi_i + h_\chi \hat \chi_1 
\hat \chi_2 \hat \chi_3 + f_0 \hat S_2 \hat \zeta_i \hat \chi_i + f_1 \hat S_1 
\hat S_1 \hat S_2 \nonumber \\ &+& f_2 \hat S_2 \hat \phi_1 \hat \phi_2 + 
{1 \over 2} f_N \hat S_1 \hat N^c_i \hat N^c_i + f_\nu \hat L_i \hat N^c_i 
\hat \phi_2 \nonumber \\ &+& f_E \hat S_1 \hat E_i \hat E^c_i + f_e \hat L_i 
\hat N^c_i \hat \phi_1 + h^e_{ijk} \hat E_i \hat e^c_j \hat \chi_k \nonumber 
\\ &+& f_U \hat S_1 \hat U_i \hat U^c_i + f_u \hat Q_i \hat U^c_i \hat \phi_2 
+ h^u_{ijk} \hat U_i \hat u^c_j \hat \chi_k \nonumber \\ &+& f_D \hat S_1 
\hat D_i \hat D^c_i + f_d \hat Q_i \hat D^c_i \hat \phi_1 + h^d_{ijk} 
\hat D_i \hat d^c_j \hat \chi_k.
\end{eqnarray}
Note that the only allowed mass term is $M_\chi$.  The heavy singlet quarks 
and leptons will have masses proportional to $\langle S_1 \rangle$ and the 
$\mu$ parameter will be proportional to $\langle S_2 \rangle$.  Because of 
the choice of $U(1)_{PQ}$, the conservation of $R$ parity is automatic as 
well.

\section{Breaking of $A_4$ and $U(1)_{PQ}$}

Consider the scalar potential consisting of the fields $\chi_i$, $\zeta_i$, 
$S_1$, and $S_2$.
\begin{eqnarray}
V &=& |M_\chi \chi_1 + h_\chi \chi_2 \chi_3 + f_0 S_2 \zeta_1|^2 + 
|M_\chi \chi_2 + h_\chi \chi_3 \chi_1 + f_0 S_2 \zeta_2|^2 \nonumber \\ 
&+& |M_\chi \chi_3 + h_\chi \chi_1 \chi_2 + f_0 S_2 \zeta_3|^2 + |f_0 S_2 
\chi_1|^2 + |f_0 S_2 \chi_2|^2 + |f_0 S_2 \chi_3|^2 \nonumber \\ 
&+& |2f_1 S_1 S_2|^2 + |f_0 (\zeta_1 \chi_1 + \zeta_2 \chi_2 + \zeta_3 
\chi_3) + f_1 S_1^2|^2.
\end{eqnarray}
This has the supersymmetric solution ($V=0$) for
\begin{equation}
u_0 = \langle \chi_1 \rangle = \langle 
\chi_2 \rangle = \langle \chi_3 \rangle = -M_\chi/h_\chi,
\end{equation}
and
\begin{equation}
v_2 = 0, ~~~ f_0 u_0 (u_1+u_2+u_3) + f_1 v_1^2 = 0,
\end{equation}
where $v_{1,2} = \langle S_{1,2} \rangle$ and $u_i = \langle \zeta_i \rangle$.

As shown in Ref.~[1], in the presence of soft supersymmetry breaking, Eq.~(8) 
will be modified to read
\begin{equation}
v_2 \sim M_{SUSY}, ~~~ f_0 u_0 (u_1+u_2+u_3) + f_1 v_1^2 \sim M_{SUSY}^2,
\end{equation}
with $v_1$ and $u_i$ of order $u_0$.  Going back to Eq.~(5), this means that 
the heavy singlet quarks and leptons will all have masses of order $M_\chi$, 
but the $\mu$ parameter is of order $M_{SUSY}$.

As shown in Ref.~[2], the quarks and charged leptons will get Dirac seesaw 
masses from their heavy counterparts, and because of the $A_4$ symmetry and 
Eq.~(7), they are diagonalized by the same unitary matrix, i.e.
\begin{equation}
U_L = {1 \over \sqrt 3} \pmatrix {1 & 1 & 1 \cr 1 & \omega & \omega^2 \cr 
1 & \omega^2 & \omega},
\end{equation}
where $\omega = e^{2 \pi i/3}$.  This means that the quark mixing matrix 
is just the identity matrix at this high scale.  It also means that the 
neutrino mass matrix is given by
\begin{equation}
{\cal M}_\nu = {f_\nu^2 \langle \phi_2 \rangle^2 \over f_N v_1} U_L^T U_L 
= {f_\nu^2 \langle \phi_2 \rangle^2 \over f_N v_1} \pmatrix {1 & 0 & 0 \cr 
0 & 0 & 1 \cr 0 & 1 & 0}.
\end{equation}
As shown in Ref.~[2], the one-loop radiative corrections of this matrix 
from the high scale to the electroweak scale automatically produce the 
phenomenologically favored neutrino mixing matrix with $\theta_{atm} = \pi/4$ 
and $\theta_{sol}$ large but less than $\pi/4$.  In the model of Ref.~[2], 
the heavy singlet fermion masses are independent of one another and unrelated 
to the $A_4$ breaking scale.  Here, they are all determined to be of order 
$M_\chi$.  Furthermore, the extra softly broken discrete $Z_3$ symmetry used 
there is no longer needed.  Both improvements are directly due to the specific 
way in which $U(1)_{PQ}$ has been realized.

The axion of this model is contained mostly in the superfield
\begin{equation}
{v_1 \hat S_1 + (2/3) (u_1+u_2+u_3) (\hat \zeta_1 + \hat \zeta_2 + \hat 
\zeta_3) \over \sqrt {v_1^2 + (4/3) (u_1 + u_2 + u_3)^2}}.
\end{equation}
However, as $S_2$ picks up a vacuum expectation value of order $M_{SUSY}$ 
and $\langle \phi_{1,2} \rangle$ become nonzero as well from electroweak 
symmetry breaking, the axion will acquire small components from each. 
Since $S_1$ couples to the heavy singlet quarks and $\phi_{1,2}$ to ordinary 
quarks, this axion is a hybrid of the two best known mechanisms \cite{ksvz,
dfsz}.

\section{Particle Spectrum}

In this model, $f_N v_1$ is the mass of the heavy singlet neutrinos and 
$[v_1^2 + (4/3)(u_1+u_2+u_3)^2]^{1/2}$ is the axion decay constant.  Both 
are presumably of order $10^{11\pm1}$ GeV to satisfy the astrophysics and 
cosmology bounds \cite{astro}.  All superfields not belonging to the MSSM 
have masses of that same order, except for the following.

\noindent (1) The very light axion of mass $\sim 10^{-4 \pm 1}$ eV which 
eliminates the strong CP problem.

\noindent (2) The corresponding physical scalar field (saxion) which has 
a mass $\sim M_{SUSY}$ but whose couplings to MSSM particles are supressed 
by $v_2/v_1 \sim 10^{-8 \pm 1}$.

\noindent (3) The corresponding supersymmetric partner (axino) which mixes 
$(\sim 10^{-9 \pm 1})$ with the neutralinos of the MSSM and may be the true 
lightest supersymmetric particle (LSP) if its mass $(\sim 2f_1 v_2)$ is 
small enough.

\noindent (4) The two superfields orthogonal to $(\hat \zeta_1 + \hat \zeta_2 
+ \hat \zeta_3)/\sqrt 3$ with scalar components of order $M_{SUSY}$ in mass 
and fermionic components of order $M_{SUSY}^2/M_\chi \sim 10^{1 \pm 1}$ keV 
in mass.  They decouple very effectively from all MSSM particles.

\section{Conclusion}

Merging two recent proposals, a comprehensive supersymmetric model (with just 
one input mass) has been presented, which results in a desirable neutrino 
mass matrix based on the discrete non-Abelian symmetry $A_4$, and removes the 
strong CP problem with a spontaneously broken $U(1)_{PQ}$ symmetry, the 
specific application of which also conserves $R$ parity and equates the $\mu$ 
parameter with $M_{SUSY}$.  The true lightest supersymmetric particle may be 
the axino of this model.

\section*{Acknowledgement}

This work was supported in part by the U.~S.~Department of Energy
under Grant No.~DE-FG03-94ER40837.

\bibliographystyle{unsrt}

\begin{thebibliography}{99}
\bibitem{ma01} E. Ma, Phys. Lett. {\bf B514}, 330 (2001).
\bibitem{bmv} K. S. Babu, E. Ma, and J. W. F. Valle, hep-ph/0206292.
\bibitem{seesaw} M. Gell-Mann, P. Ramond, and R. Slansky, in 
{\it Supergravity}, edited by P. van Nieuwenhuizen and D. Z. Freedman 
(North-Holland, Amsterdam, 1979), p.~315; T. Yanagida, in {\it Proceedings 
of the Workshop on the Unified Theory and the Baryon Number in the Universe}, 
edited by O. Sawada and A. Sugamoto (KEK, Tsukuba, Japan, 1979), p.~95; 
R. N. Mohapatra and G. Senjanovic, Phys. Rev. Lett. {\bf 44}, 912 (1980).
\bibitem{pq} R. D. Peccei and H. R. Quinn, Phys. Rev. Lett. {\bf 38}, 1440 
(1977).
\bibitem{qcdcp} C. G. Callan, R. F. Dashen, and D. J. Gross, Phys. Lett. 
{\bf B63}, 334 (1976); R. Jackiw and C. Rebbi, Phys. Rev. Lett. {\bf 37}, 
172 (1976). 
\bibitem{axi} S. Weinberg, Phys. Rev. Lett. {\bf 40}, 223 (1978); F. Wilczek, 
Phys. Rev. Lett. {\bf 40}, 279 (1978).
\bibitem{a4} E. Ma and G. Rajasekaran, Phys. Rev. {\bf D64}, 113012 (2001); 
E. Ma, Mod. Phys. Lett. {\bf A17}, 289 (2002); E. Ma, Mod. Phys. Lett. 
{\bf A17}, 627 (2002).
\bibitem{ksvz} J. E. Kim, Phys. Rev. Lett. {\bf 43}, 103 (1979); M. A. 
Shifman, A. I. Vainshtein, and V. I. Zakharov, Nucl. Phys. {\bf B166}, 493 
(1980).
\bibitem{dfsz} M. Dine, W. Fischler, and M. Srednicki, Phys. Lett. {\bf B104}, 
199 (1981); A. R. Zhitnitsky, Sov. J. Nucl. Phys. {\bf 31}, 260 (1980).
\bibitem{astro} G. G. Raffelt, Ann. Rev. Nucl. Part. Sci. {\bf 49}, 163 (1999).

\end{thebibliography}

\end{document}